\documentclass{JHEP3}
\usepackage{graphicx,amsmath,amssymb}
\usepackage{epsfig,multicol}
\usepackage{epsf}
\usepackage{epstopdf}
\input{epsf.sty}

\usepackage{graphicx}
\usepackage{amssymb}
\usepackage{amsmath}

\def\p{\partial}

\def\half{{1\over 2}}
\def\({\left(}
\def\){\right)}
\def\[{\left[}
\def\]{\right]}

\def\e{\begin{equation}}
\def\q{\end{equation}}
\def\m{\begin{eqnarray}}
\def\n{\end{eqnarray}}

\title{Consistency relation for the Lorentz invariant single-field inflation}
\author{Qing-Guo Huang \footnote{huangqg@itp.ac.cn}
\\\small{\em
Key Laboratory of Frontiers in Theoretical Physics,
Institute of Theoretical Physics, Chinese Academy
of Sciences, Beijing 100190, China}
\\\small{\em School of Physics, Korea Institute for Advanced Study,
207-43, Cheongryangri-Dong, Dongdaemun-Gu, Seoul 130-722, Korea } }

\abstract{
In this paper we compute the sizes of
equilateral and orthogonal shape bispectrum for the general
Lorentz invariant single-field inflation. The stability of
field theory implies a non-negative square of sound speed
which leads to a consistency relation between the sizes of
orthogonal and equilateral shape bispectrum, namely
$f_{NL}^{orth.}\leq -0.054 f_{NL}^{equil.}$. In particular,
for the single-field Dirac-Born-Infeld (DBI) inflation, the
consistency relation becomes $f_{NL}^{orth.}= 0.070
f_{NL}^{equil.}\leq 0$. These consistency relations are
also valid in the mixed scenario where the quantum
fluctuations of some other light scalar fields contribute
to a part of total curvature perturbation on the
super-horizon scale and may generate a local form
bispectrum.  A distinguishing prediction of the mixed
scenario is $\tau_{NL}^{loc.}>({6\over 5}f_{NL}^{loc.})^2$.
Comparing these consistency relations to WMAP 7yr data,
there is still a big room for the Lorentz invariant
inflation, but DBI inflation has been disfavored at more
than $68\%$ CL.
}


\begin{document}

\section{Introduction}

The quantum fluctuation during inflation seeds the temperature anisotropies in the cosmic microwave background radiation (CMBR) and formation of large-scale structure of galaxies in our universe. The primordial cosmological perturbations are so tiny that the generation and evolution of fluctuations has been investigated within linear perturbation theory. Within this approach, the primordial perturbation is Gaussian; or equivalently, its Fourier components are uncorrelated and have random phases. The simplest version of inflation predicts such a nearly Gaussian distribution \cite{Maldacena:2002vr}. A non-vanishing three-point correlation function of the curvature perturbation, or its Fourier transform, the bispectrum, is an important indicator of a non-Gaussian feature in the cosmological perturbations because it represents the lowest order statistics able to distinguish non-Gaussian from Gaussian perturbations. Recently the non-Gaussianity emerges as a more and more important observable.

In general the bispectrum depends on configuration in momentum space \cite{Babich:2004gb,Senatore:2009gt}.
A large non-local shape, such as equilateral and orthogonal shape, bispectrum is obtained in the single-field inflation model where the higher derivative terms are involved \cite{Garriga:1999vw,Alishahiha:2004eh,Bean:2007hc,Lidsey:2007gq,Peiris:2007gz,Bean:2007eh,Chen:2006nt,ArkaniHamed:2003uz}. In the multi-field inflation the perturbations along the entropy directions which are transverse to the motion direction (adiabatic direction) can be converted into the adiabatic perturbation on the super-horizon scale and generate a local form bispectrum, for example the curvature perturbation generated at the end of multi-field inflation due to the curved geometry of the hyper-surface for inflation to end \cite{Lyth:2005qk,Sasaki:2008uc,Huang:2009vk} and the curvaton model \cite{ Enqvist:2001zp,Lyth:2001nq,Moroi:2001ct,Sasaki:2006kq,Huang:2008zj}.

In this paper we mainly focus on equilateral, orthogonal, and local form bispectrum whose sizes are measured by $f_{NL}^{equil.}$, $f_{NL}^{orth.}$ and $f_{NL}^{loc.}$ respectively.
WMAP 7yr data \cite{Komatsu:2010fb} indicates
\e
f_{NL}^{loc.}=32\pm 21,\quad f_{NL}^{equil.}=26\pm 140,\quad f_{NL}^{orth.}=-202\pm 104
\q
at $68\%$ CL; and
\e
-10<f_{NL}^{loc.}<74,\quad -214<f_{NL}^{equil.}<266,\quad -410<f_{NL}^{orth.}<6
\q
at $95\%$ CL. Up to now the data implies that the distribution of the primordial curvature perturbation deviates from an exact Gaussian distribution at $1 \sigma$ level, but a Gaussian distribution is still consistent with the data at $2 \sigma$ level.

In last decades a lot of efforts have been put for
constructing a realistic inflation model from string
theory. Many of these models fall into a class called brane
inflation \cite{Dvali:1998pa}. The dynamics of brane is
governed by the DBI action. In Sec.~2 we will figure out
the consistency relation between the sizes of equilateral
and orthogonal shape bispectrum for the general Lorentz
invariant single-field inflation model, in particular for
the DBI inflation. In
order to achieve a bispectrum which is large not only for
the non-local shape but also for the local shape, we
suggest a mixed scenario where the entropy fluctuations are
assumed to be converted into a part of total curvature
perturbation and produce a local form bispectrum at/after
the end of inflation in Sec.~3. We find that the
consistency relations derived in Sec.~2 are
also valid in this mixed scenario. Comparing to the WMAP
7yr data, the single-field DBI inflation has been disfavored at more
than $1\sigma$ level. A short summary is included in Sec.~4.

\section{Consistency relations}

The CMB temperature anisotropy in the Sachs-Wolfe limit is given by $\Delta T/T=-{1\over 3}\Phi$, where $\Phi$ is the Bardeen's curvature perturbation. The power spectrum of the curvature perturbation is defined by
\e
\langle \Phi_{{\bf k}_1}\Phi_{{\bf k}_2}\rangle=(2\pi)^3\delta^{(3)}({\bf k}_1+{\bf k}_2){\Delta_\Phi\over k_1^s},
\q
where
\e
s=4-n_s,
\q
and $n_s$ is so-called spectral index.
A general description of non-Gaussianity at the leading order is the bispectrum of curvature perturbation
\m
\langle \Phi_{{\bf k}_1} \Phi_{{\bf k}_2} \Phi_{{\bf k}_3} \rangle= (2\pi)^3\delta^{(3)} (\sum_{i=1}^3 {\bf k}_i) F(k_1,k_2,k_3),
\label{bis}
\n
where  $\Phi_{\bf k}$ is the Fourier mode of curvature perturbation in momentum space and $F(k_1,k_2,k_3)$ depends on the configuration in momentum space. For the local, equilateral, and orthogonal form bispectrum, $F(k_1,k_2,k_3)$ are respectively given by
\m
F^{loc.}(k_1,k_2,k_3)&=&\Delta_\Phi^2 f_{NL}^{loc.} {\tilde F}^{loc.}(k_1,k_2,k_3), \\
F^{equil.}(k_1,k_2,k_3)&=&\Delta_\Phi^2 f_{NL}^{equil.} {\tilde F}^{equil.}(k_1,k_2,k_3), \\
F^{orth.}(k_1,k_2,k_3)&=&\Delta_\Phi^2 f_{NL}^{orth.} {\tilde F}^{orth.}(k_1,k_2,k_3),
\n
where
\m
{\tilde F}^{loc.}(k_1,k_2,k_3)&=&2 \[{1\over k_1^s k_2^s}+(2\ \hbox{perm.})\], \\
{\tilde F}^{equil.}(k_1,k_2,k_3)&=&6 \[ -{1\over k_1^sk_2^s}-{1\over k_2^sk_3^s}-{1\over k_3^sk_1^s} \right. \nonumber \\
&& \left. -{2\over (k_1k_2k_3)^{2s/3}} +\({1\over k_1^{s/3} k_2^{2s/3}k_3^s}+(5\ \hbox{perm.})\) \right], \\
{\tilde F}^{orth.}(k_1,k_2,k_3)&=&6 \[ -{3\over k_1^sk_2^s}-{3\over k_2^sk_3^s}-{3\over k_3^sk_1^s} \right. \nonumber \\
&& \left. -{8\over (k_1k_2k_3)^{2s/3}} +\({3\over k_1^{s/3} k_2^{2s/3}k_3^s}+(5\ \hbox{perm.})\) \right].
\n
These three forms are nearly orthogonal to one another. They probe different aspects of the physics of inflation.

First of all, let's focus on the general single-field inflation.
In \cite{Chen:2006nt} the bispectrum was calculated in the single-field inflation model with action
\e
S=\int d^4x\sqrt{-g}\[{M_p^2\over 2}R+P(X,\phi)\],
\q
where $X=-\half g^{\mu\nu}\p_\mu\phi\p_\nu\phi$. This action is the most general Lorentz invariant action for inflaton $\phi$ minimally coupled to Einstein gravity.
For the case with large bispectrum ($c_s\ll 1$ and/or $\lambda/\Sigma\gg 1$),
\m
F(k_1,k_2,k_3)= \Delta_\Phi^2 {\cal B}(k_1,k_2,k_3),
\n
where
\m
{\cal B}(k_1,k_2,k_3)&=&10\({1\over c_s^2}-1-{2\lambda \over \Sigma}\) {1\over k_1 k_2 k_3 K^3} \nonumber \\
&+&{20\over 3}\({1\over c_s^2}-1\) {1\over k_1^3k_2^3k_3^3}\(-{1\over K}\sum_{i>j}k_i^2k_j^2 \right. \nonumber \\
&+&\left. {1\over 2K^2}\sum_{i\neq j}k_i^2k_j^3 + {1\over 8}\sum_i k_i^3\),
\n
up to ${\cal{O}} (\epsilon,\eta,{\dot c_s}/(c_sH))$,
and
\m
K&=&k_1+k_2+k_3, \\
c_s^2&=&{P_{,X}\over P_{,X}+2XP_{,XX}},\\
\lambda &=& X^2 P_{,XX} + {2\over 3}X^3P_{,XXX},\\
\Sigma &=& XP_{,X}+2X^2P_{,XX}.
\n
Evaluating $F(k_1,k_2,k_3)$ at the equilateral triangle
$k_1=k_2=k_3\equiv k$, we obtain
\m
f_{NL}^{1/c_s^2}&=&-{85\over 324}\({1\over c_s^2}-1\), \label{fcs}\\
f_{NL}^{\lambda/\Sigma}&=&-{10\over 81}{\lambda\over \Sigma}, \label{fls}
\n
where $f_{NL}$ is defined by
\e
F(k,k,k)=f_{NL}\cdot {6\Delta_\Phi^2\over k^6}.
\q
In the literatures, ones take $f_{NL}^{1/c_s^2}$ and
$f_{NL}^{\lambda/\Sigma}$ as the sizes of the equilateral
shape bispectrum generated by the terms with $1/c_s^2$ and
$\lambda/\Sigma$ respectively. However, we need to stress
that this simple convention is not the same as that adopted
by WMAP group. In order to compare to WMAP results, we need to
project the bispectrum in general Lorentz invariant
single-field inflation to the different templates and work
out the corresponding non-Gaussianity parameters.
Following the definition of a 3-dimensional scalar product
between the shapes $F_{(1)}$ and $F_{(2)}$ in
\cite{Babich:2004gb,Senatore:2009gt}
\e
F_{(1)}\cdot F_{(2)}=\sum_{k_i^{physical}} {F_{(1)}(k_1,k_2,k_3)F_{(2)}(k_1,k_2,k_3)\over P_{k_1}P_{k_2}P_{k_3}},
\q
where $P_{k}$ represents the power spectrum and
$k_i^{physical}$ means that only the ${\bf k}$'s that form
a triangle are included.
The sizes of equilateral and orthogonal shape bispectrum
are defined by
\m
f_{NL}^{equil.}&=&\({F_{\lambda/\Sigma=0}\cdot F^{equil.}\over F^{equil.}\cdot F^{equil.}}\)_{f_{NL}\Delta_\Phi^2=1} f_{NL}^{1/c_s^2}+\({F_{c_s=1}\cdot F^{equil.}\over F^{equil.}\cdot F^{equil.}}\)_{f_{NL}\Delta_\Phi^2=1} f_{NL}^{\lambda/\Sigma},\\
f_{NL}^{equil.}&=&\({F_{\lambda/\Sigma=0}\cdot F^{orth.}\over F^{orth.}\cdot F^{orth.}}\)_{f_{NL}\Delta_\Phi^2=1} f_{NL}^{1/c_s^2}+\({F_{c_s=1}\cdot F^{orth.}\over F^{orth.}\cdot F^{orth.}}\)_{f_{NL}\Delta_\Phi^2=1} f_{NL}^{\lambda/\Sigma}.
\n
Therefore we obtain
\m
f_{NL}^{equil.}&=& -0.2728\({1\over c_s^2}-1\)-0.1494{\lambda\over \Sigma},\\
f_{NL}^{orth.}&=& -0.02831\({1\over c_s^2}-1\)+0.008114{\lambda\over \Sigma}.
\n
Similarly, $f_{NL}^{loc.}=0$ which implies the Lorentz
invariant single-field inflation cannot generate a local
form bispectrum.

The parameters $c_s^2$ and $\lambda/\Sigma$ can be fixed once $f_{NL}^{equil.}$ and $f_{NL}^{orth.}$ are detected,
\m
{1\over c_s^2}-1&=& -1.260 f_{NL}^{equil.}-23.19 f_{NL}^{orth.}, \\
{\lambda \over \Sigma}&=& -4.394 f_{NL}^{equil.}+42.34 f_{NL}^{orth.}.
\n
Stability of the field theory implies $c_s^2\geq 0$ which leads to
\e
f_{NL}^{orth.}\leq -0.054 f_{NL}^{equil.}.
\label{cgb}
\q
It is the consistency condition for the general single-field inflation without breaking Lorentz symmetry.

Nowadays string theory is supposed to be the only one self-consistent theory of quantum gravity. A very popular inflation model in string theory is brane inflation \cite{Dvali:1998pa}. In KKLT \cite{Kachru:2003aw} set up, brane can move very fast (close to the speed of light) in the Klebanov-Strassler (KS) throat, and the full DBI action for the inflaton field should be taken into account. For DBI inflation \cite{ Alishahiha:2004eh,Bean:2007eh}, the action of $\phi$ takes the form
\e
P=-f^{-1}(\phi)\sqrt{1-2Xf(\phi)}+f^{-1}(\phi)-V(\phi).
\q
Therefore
\m
c_s&=&\sqrt{1-2Xf(\phi)}, \\
{\lambda\over \Sigma}&=&\half \({1\over c_s^2}-1\).
\n
The sizes of equilateral and orthogonal form bispectrum are given by
\m
f_{NL}^{equil.}&=& -0.3475\({1\over c_s^2}-1\),\\
f_{NL}^{orth.}&=& -0.02425\({1\over c_s^2}-1\).
\n
Because the radial velocity of brane in the KS throat is limited by the speed of light, we have $c_s^2\geq 0$ which is nothing but the condition for the stability of field theory. Therefore we obtain
\e
f_{NL}^{orth.}= 0.070 f_{NL}^{equil.}\leq 0.
\label{cdb}
\q

\begin{figure}[h]
\begin{center}
\includegraphics[width=10cm]{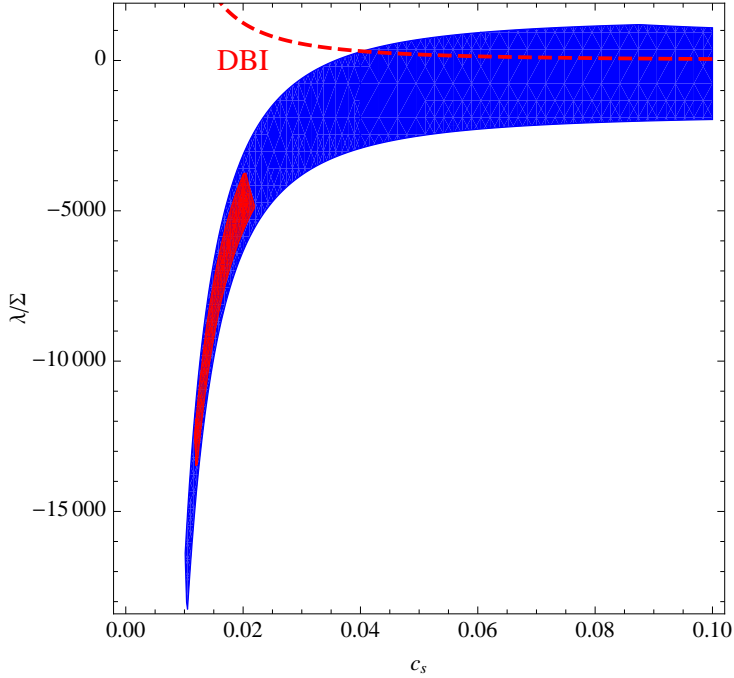}
\end{center}
\caption{The allowed parameter space for the general Lorentz invariant single-field inflation. The red and blue patches are the allowed regions at $68\%$ and $95\%$ CL respectively. The red dashed line corresponds to the prediction of DBI inflation. }
\label{fig:fitting}
\end{figure}

As we know, the single-field inflation can only generate non-local form non-Gaussianity. Comparing to the WMAP 7yr data (constraints on $f_{NL}^{equil.}$ and $f_{NL}^{orth.}$), the allowed region for the general Lorentz invariant single-field inflation shows up in Fig. \ref{fig:fitting}.
We see that the single-field DBI inflation has been disfavored at more
than $1\sigma$ level. \footnote{The so-called UV model of
single-field DBI inflation has been known to be
incompatible to the observations in
\cite{Bean:2007hc,Lidsey:2007gq,Peiris:2007gz,Bean:2007eh}.
Here we quote the constraints on the non-Gaussianity
parameters from WMAP group. The unresolved
extra-galactic point sources might bias these constraints. }

\section{Mixed scenario}

However, a convincing detection of a large local form bispectrum will rule out all single-field inflation models (not only the slow-roll model). If one wants to obtain a bispectrum which is large not only for the non-local shape but also for the local shape, one needs a mixed scenario like that for the curvaton model proposed in \cite{Huang:2008zj}. Here we give a more general discussion about the bispectrum in the multi-field inflation where the trajectory of inflaton fields is assumed to be a straight line in the field space during inflation. Denote the scalar field $\phi_s$ to be transverse to the inflaton field $\phi$ along the adiabatic direction. In the mixed scenario, both quantum fluctuations of $\phi$ and $\phi_s$ are assumed to contribute to the total curvature perturbation which seeds the temperature anisotropies in the CMBR. For simplicity, $\phi_s$ is decoupled to $\phi$, namely $\langle \delta\phi_{{\bf k}_1}, \delta\phi_{s,{{\bf k}_2}}\rangle=0$. Considering
\e
\Phi=\Phi^\phi+\Phi^{\phi_s},
\q
and $\langle \Phi^\phi_{{\bf k}_1}, \Phi^{\phi_s}_{{\bf k}_2}\rangle=0$, we obtain
\e
\Delta_\Phi^{tot.}=\Delta_\Phi^\phi+\Delta_\Phi^{\phi_s}.
\q
For convenience, we introduce a new parameter $\beta$ which is defined by
\e
\beta=\Delta_\Phi^{\phi_s}/\Delta_\Phi^{tot.}.
\q
Here $\beta\in [0,1]$, and $\Delta_\Phi^\phi=(1-\beta)\Delta_\Phi^{tot.}$. The index of power spectrum becomes
\e
n_s^{tot.}=(1-\beta)n_s^\phi+\beta n_s^{\phi_s}.
\q
Since $\phi_s$ is orthogonal to the adiabatic direction and
hence the fluctuation of $\phi_s$ only perturbs the value
of $\phi_s$, not the energy density, it does not generate
curvature perturbation during inflation, but its
fluctuation can be converted into the adiabatic one
at/after the end of inflation on the super-horizon scale
\cite{Lyth:2005qk,Sasaki:2008uc,Huang:2009vk,Enqvist:2001zp,Lyth:2001nq,Moroi:2001ct,Sasaki:2006kq,Huang:2008zj}.
So the non-Gaussianity caused by the fluctuation of
$\phi_s$ has a local form. \footnote{If the field
becomes non-Gaussian at the horizon exit, the entropy
fluctuations may contribute a non-local shape bispectrum as
well. We ignore this case in this paper. } Denoting that the
size of local form bispectrum generated by the quantum
fluctuation of $\phi_s$ as $f_{NL}^{\phi_s}$, we obtain
\m
F(k_1,k_2,k_3)&=&(\Delta_\Phi^{\phi_s})^2f_{NL}^{\phi_s}{\tilde F}^{loc.}(k_1,k_2,k_3) \nonumber \\
&+&(\Delta_\Phi^{\phi})^2 {\cal B}(k_1,k_2,k_3).
\n
Considering $\Delta_\Phi^{\phi_s}=\beta\Delta_\Phi^{tot.}$ and $\Delta_\Phi^\phi=(1-\beta)\Delta_\Phi^{tot.}$, the effective sizes of local, equilateral and orthogonal form bispectrum are respectively given by
\m
f_{NL}^{loc.}&=&\beta^2 f_{NL}^{\phi_s},\\
f_{NL}^{equil.}&=&(1-\beta)^2 f_{NL}^{\phi, equil.},\\
f_{NL}^{orth.}&=&(1-\beta)^2 f_{NL}^{\phi, orth.}.
\n
Since $f_{NL}^{equil.}$ and $f_{NL}^{orth.}$ are rescaled by a same factor, the consistency relations (\ref{cgb}) and (\ref{cdb}) are still valid in the mixed scenario.

Because $\beta$ is a free parameter, it is better to use the
consistency relations to constrain the inflation model. In the mixed scenario, the coordinates in Fig.~(\ref{fig:fitting}) should be replaced by $(1-\beta)\lambda/\Sigma$ and $c_s/(1-\beta)$. Comparing to
the bounds on $f_{NL}^{equil.}$ and $f_{NL}^{orth.}$ from WMAP 7yr
data, we find that the single-field DBI inflation is disfavored at more than $68\%$
CL. This conclusion is valid even in the mixed scenario for
producing a local form bispectrum, either.

In \cite{Huang:2008zj} we first pointed out the enhancement of $\tau_{NL}^{loc.}$ comparing to $({6\over 5}f_{NL}^{loc.})^2$ in the mixed curvaton model. Actually it is a generic result. The curvature perturbation generated by $\phi_s$ on the super-horizon scale can be expanded as follows
\e
\Phi^{\phi_s}=\Phi_L^{\phi_s}+f_{NL}^{\phi_s} \(\Phi_L^{\phi_s}\)^2+g_{NL}^{\phi_s} \(\Phi_L^{\phi_s}\)^3+...\ ,
\q
where $\Phi_L^{\phi_s}$ is the linear part of the curvature perturbation.
Comparing to the definition of $\tau_{NL}^{loc.}$
\m
&&\langle \Phi_{{\bf k}_1} \Phi_{{\bf k}_2} \Phi_{{\bf k}_3} \Phi_{{\bf k}_4} \rangle \label{gnl}  \\
&=& (2\pi)^3 \delta^{(3)}(\sum_{i=1}^4{\bf k}_i) \cdot \[ 6g_{NL}^{loc.}\Delta_\Phi^3 \cdot {\sum_{i=1}^4 k_i^s\over \prod_{i=1}^4 k_i^s} \right. \nonumber \\
&+& \left. {25\over 18} \tau_{NL}^{loc.} \Delta_\Phi^3 \cdot \({1 \over k_{12}^s k_2^s k_3^s}+23 \ \hbox{perms}.\)\]\ , \nonumber
\n
we find
\e
\tau_{NL}^{loc.}= \beta^3 ({6\over 5}f_{NL}^{\phi_s})^2.
\q
Here we consider that the quantum fluctuation of $\phi$ does not generated local form bispectrum and trispectrum.
Since $f_{NL}^{loc.}= \beta^2 f_{NL}^{\phi_s}$,
\e
\tau_{NL}^{loc.}= {1\over \beta}({6\over 5}f_{NL}^{loc.})^2
\q
which is enhanced by a factor $1/\beta$. If both local and non-local shape bispectrum are detected by the upcoming cosmological observations, such as Planck satellite, the mixed scenario must be called for, and $\beta$ should be smaller than one. Therefore $\tau_{NL}^{loc.}>({6\over 5}f_{NL}^{loc.})$. This is a distinguishing prediction. As we know, a local form trispectrum with $\tau_{NL}^{loc.}>560$ can be detected by Planck at $2\sigma$ level \cite{Kogo:2006kh}. It is very exciting to check the consistency relation between $\tau_{NL}^{loc.}$ and $f_{NL}^{loc.}$ and fix the value of $\beta$ from experiments in the near future.

In addition, one may worry that the isocurvature perturbations in the mixed scenario might be too big to fit the WMAP data \cite{Komatsu:2010fb}. However, whether the isocurvature perturbations are generated in the multi-field inflation depends on the detail of reheating. For example, if the cold dark matter is not the direct decay product of the adiabatic and entropic fields and the cold dark matter is generated after all of these fields decay completely, the perturbations from multi fields do not generate the detectable isocurvature perturbations, and the mixed scenario is free from the constraint on the isocurvature perturbations.

\section{Conclusion}

To summarize, the stability of the field theory leads to a
consistency relation (\ref{cgb}) between the equilateral
and orthogonal shape bispectrum in the general Lorentz
invariant single-field inflation. Even though there is a
big room for the general Lorentz invariant single-field
inflation, some well-known Lorentz invariant single-field
inflation models, such as DBI inflation, have been tightly
constrained by the WMAP 7yr data. The consistency relations
(\ref{cgb}) and (\ref{cdb}) are also valid even in the
mixed scenario where the fluctuations along the entropy
directions also make contributions to the total curvature
perturbation on the super-horizon scale and generate a
local form bispectrum and trispectrum.  We conclude that
the single-field DBI inflation model is disfavored by WMAP 7yr
data at more than $68\%$ CL. So the inflation driven by a
D-brane in string theory seems unlikely, but a multi-field
version might be used to evade some of the observational
constraints. Similarly, the consistency relation for the
power-law K-inflation \cite{ArmendarizPicon:1999rj} is
$f_{NL}^{orth.}\simeq 0.1 f_{NL}^{equil.}<0$ which is also
disfavored.

As we know, the mixed scenario provides the only mechanism to achieve an inflation model with not only a large local form but also large non-local shape bispectrum and trispectrum. A distinguishing prediction of this scenario is $\tau_{NL}^{loc.}>({6\over 5}f_{NL}^{loc})^2$.

\noindent {\bf Acknowledgments}

We would like to thank P.~Chingangbam and X.~Gao for useful discussions. This work is supported by the project of Knowledge Innovation Program of Chinese Academy of Science.





\newpage

\begin{thebibliography}{99}


\bibitem{Maldacena:2002vr}
  J.~M.~Maldacena,
  ``Non-Gaussian features of primordial fluctuations in single field
  inflationary models,''
  JHEP {\bf 0305}, 013 (2003)
  [arXiv:astro-ph/0210603].

\bibitem{Babich:2004gb}
  D.~Babich, P.~Creminelli and M.~Zaldarriaga,
  ``The shape of non-Gaussianities,''
  JCAP {\bf 0408}, 009 (2004)
  [arXiv:astro-ph/0405356].


\bibitem{Senatore:2009gt}
  L.~Senatore, K.~M.~Smith and M.~Zaldarriaga,
  ``Non-Gaussianities in Single Field Inflation and their Optimal Limits from
  the WMAP 5-year Data,''
  JCAP {\bf 1001}, 028 (2010)
  [arXiv:0905.3746 [astro-ph.CO]].

\bibitem{Garriga:1999vw}
  J.~Garriga and V.~F.~Mukhanov,
  ``Perturbations in k-inflation,''
  Phys.\ Lett.\  B {\bf 458}, 219 (1999)
  [arXiv:hep-th/9904176].


\bibitem{Alishahiha:2004eh}
  M.~Alishahiha, E.~Silverstein and D.~Tong,
  ``DBI in the sky,''
  Phys.\ Rev.\  D {\bf 70}, 123505 (2004)
  [arXiv:hep-th/0404084].

\bibitem{Bean:2007hc}
  R.~Bean, S.~E.~Shandera, S.~H.~Henry Tye and J.~Xu,
  ``Comparing Brane Inflation to WMAP,''
  JCAP {\bf 0705}, 004 (2007)
  [arXiv:hep-th/0702107].

\bibitem{Lidsey:2007gq}
  J.~E.~Lidsey and I.~Huston,
  ``Gravitational wave constraints on Dirac-Born-Infeld inflation,''
  JCAP {\bf 0707}, 002 (2007)
  [arXiv:0705.0240 [hep-th]].

\bibitem{Peiris:2007gz}
  H.~V.~Peiris, D.~Baumann, B.~Friedman and A.~Cooray,
  ``Phenomenology of D-Brane Inflation with General Speed of Sound,''
  Phys.\ Rev.\  D {\bf 76}, 103517 (2007)
  [arXiv:0706.1240 [astro-ph]].


\bibitem{Bean:2007eh}
  R.~Bean, X.~Chen, H.~Peiris and J.~Xu,
  ``Comparing Infrared Dirac-Born-Infeld Brane Inflation to Observations,''
  Phys.\ Rev.\  D {\bf 77}, 023527 (2008)
  [arXiv:0710.1812 [hep-th]].

\bibitem{Chen:2006nt}
  X.~Chen, M.~x.~Huang, S.~Kachru and G.~Shiu,
  ``Observational signatures and non-Gaussianities of general single field inflation,''
  JCAP {\bf 0701}, 002 (2007)
  [arXiv:hep-th/0605045].


\bibitem{ArkaniHamed:2003uz}
  N.~Arkani-Hamed, P.~Creminelli, S.~Mukohyama and M.~Zaldarriaga,
  ``Ghost Inflation,''
  JCAP {\bf 0404}, 001 (2004)
  [arXiv:hep-th/0312100].


\bibitem{Lyth:2005qk}
  D.~H.~Lyth,
  ``Generating the curvature perturbation at the end of inflation,''
  JCAP {\bf 0511}, 006 (2005)
  [arXiv:astro-ph/0510443].


\bibitem{Sasaki:2008uc}
  M.~Sasaki,
  ``Multi-brid inflation and non-Gaussianity,''
  Prog.\ Theor.\ Phys.\  {\bf 120}, 159 (2008)
  [arXiv:0805.0974 [astro-ph]].


\bibitem{Huang:2009vk}
  Q.~G.~Huang,
  ``A geometric description of the non-Gaussianity generated at the end of
  multi-field inflation,''
  JCAP {\bf 0906}, 035 (2009)
  [arXiv:0904.2649 [hep-th]].


\bibitem{Enqvist:2001zp}
  K.~Enqvist and M.~S.~Sloth,
  ``Adiabatic CMB perturbations in pre big bang string cosmology,''
  Nucl.\ Phys.\  B {\bf 626}, 395 (2002)
  [arXiv:hep-ph/0109214].

\bibitem{Lyth:2001nq}
  D.~H.~Lyth and D.~Wands,
  ``Generating the curvature perturbation without an inflaton,''
  Phys.\ Lett.\  B {\bf 524}, 5 (2002)
  [arXiv:hep-ph/0110002].

\bibitem{Moroi:2001ct}
  T.~Moroi and T.~Takahashi,
  ``Effects of cosmological moduli fields on cosmic microwave background,''
  Phys.\ Lett.\  B {\bf 522}, 215 (2001)
  [Erratum-ibid.\  B {\bf 539}, 303 (2002)]
  [arXiv:hep-ph/0110096].

\bibitem{Sasaki:2006kq}
  M.~Sasaki, J.~Valiviita and D.~Wands,
  ``Non-gaussianity of the primordial perturbation in the curvaton model,''
  Phys.\ Rev.\  D {\bf 74}, 103003 (2006)
  [arXiv:astro-ph/0607627].

\bibitem{Huang:2008zj}
  Q.~G.~Huang,
  ``A Curvaton with a Polynomial Potential,''
  JCAP {\bf 0811}, 005 (2008)
  [arXiv:0808.1793 [hep-th]].




\bibitem{Komatsu:2010fb}
  E.~Komatsu {\it et al.},
  ``Seven-Year Wilkinson Microwave Anisotropy Probe (WMAP) Observations:
  Cosmological Interpretation,''
  arXiv:1001.4538 [astro-ph.CO].

\bibitem{Dvali:1998pa}
  G.~R.~Dvali and S.~H.~H.~Tye,
  ``Brane inflation,''
  Phys.\ Lett.\  B {\bf 450}, 72 (1999)
  [arXiv:hep-ph/9812483].


\bibitem{Kachru:2003aw}
  S.~Kachru, R.~Kallosh, A.~D.~Linde and S.~P.~Trivedi,
  ``De Sitter vacua in string theory,''
  Phys.\ Rev.\  D {\bf 68}, 046005 (2003)
  [arXiv:hep-th/0301240].




\bibitem{Kogo:2006kh}
  N.~Kogo and E.~Komatsu,
 ``Angular Trispectrum of CMB Temperature Anisotropy from Primordial
  Non-Gaussianity with the Full Radiation Transfer Function,''
  Phys.\ Rev.\  D {\bf 73}, 083007 (2006)
  [arXiv:astro-ph/0602099].


\bibitem{ArmendarizPicon:1999rj}
  C.~Armendariz-Picon, T.~Damour and V.~F.~Mukhanov,
  ``k-Inflation,''
  Phys.\ Lett.\  B {\bf 458}, 209 (1999)
  [arXiv:hep-th/9904075].




\end{thebibliography}
\end{document}